\documentclass[acs,jpcl,amsmath,amssymb,superscriptaddress,reprint]{revtex4-1}

\usepackage[utf8]{inputenc}
\usepackage{graphicx}
\usepackage{dcolumn}
\usepackage{bm}
\usepackage{blindtext}
\usepackage{physics}
\usepackage[justification=justified, format=plain]{caption}
\newcommand\Tstrut{\rule{0pt}{2.9ex}}         
\newcommand\Bstrut{\rule[-1.2ex]{0pt}{0pt}}   
\newcommand\TBstrut{\Tstrut\Bstrut}           

\usepackage{xspace}
\usepackage{siunitx}
\usepackage{tikz}
\usepackage{ulem}
\usetikzlibrary{shapes,arrows}

\begin{document}


\title[]{Nuclear Quantum Effects in liquid water at near classical computational cost using the adaptive Quantum Thermal Bath }
\author{Nastasia Mauger$^\#$}
\affiliation{Sorbonne Universit\'e, LCT, UMR 7616 CNRS, F-75005, Paris, France}
\author{Thomas Plé$^\#$}
\affiliation{CNRS, Sorbonne Universit\'e, Institut des NanoSciences de Paris,
UMR 7588, 4 Place Jussieu, F-75005 Paris, France}
\author{Louis Lagardère*}
\affiliation{Sorbonne Universit\'e, LCT, UMR 7616 CNRS, F-75005, Paris, France}
\author{Sara Bonella}
\affiliation{CECAM Centre Européen de Calcul Atomique et Moléculaire, École Polytechnique Fédérale de Lausanne, Batochimie, Avenue Forel 2, 1015 Lausanne, Switzerland\\
$^{\#}$~These authors contributed equally to this work}
\author{\'Etienne Mangaud}
\affiliation{CNRS, Sorbonne Universit\'e, Institut des NanoSciences de Paris,
UMR 7588, 4 Place Jussieu, F-75005 Paris, France}
\author{Jean-Philip Piquemal*}
\affiliation{Sorbonne Universit\'e, LCT, UMR 7616 CNRS, F-75005, Paris, France}
 \altaffiliation[Also at ]{Institut Universitaire de France, 75005, Paris , France}
 \altaffiliation[Also at ]{Department of Biomedical Engineering, The University of Texas at Austin, USA}
\email{jean-philip.piquemal@sorbonne-universite.fr}
\author{Simon Huppert*}
\affiliation{CNRS, Sorbonne Universit\'e, Institut des NanoSciences de Paris,
UMR 7588, 4 Place Jussieu, F-75005 Paris, France}

\date{\today}

\begin{abstract}
We demonstrate the accuracy and efficiency of a recently introduced approach to account for nuclear quantum effects (NQE) in molecular simulations: the adaptive Quantum Thermal Bath (adQTB). In this method, zero point energy is introduced through a generalized Langevin thermostat designed to precisely enforce the quantum fluctuation-dissipation theorem. We propose a refined adQTB algorithm with improved accuracy and we report adQTB simulations of liquid water. Through extensive comparison with reference path integral calculations, we demonstrate that it provides excellent accuracy for a broad range of structural and thermodynamic observables as well as infrared vibrational spectra. The adQTB has a computational cost comparable to classical molecular dynamics, enabling simulations of up to millions of degrees of freedom.
\end{abstract}

\maketitle

Nuclear quantum effects play a major role in a wide range of physical and chemical processes where light atoms, and especially hydrogen, are involved\cite{benoit1998tunnelling, miura1998ab, rossi2016anharmonic, monacelli2020black}. In particular, a few studies point to their importance in biological systems\cite{agarwal2002network,perez2010enol,wang2014quantum}, where hydrogen-bonding is ubiquitous, but realistic atomic-scale simulations in that area remain scarce. For such large and complex systems, the most common approach has been to include NQEs implicitly, by fitting analytical potential energy surface models in order to recover experimental thermodynamic properties when performing simulations with classical nuclei\cite{paesanireview,onufrievreview}. This strategy potentially limits transferability and its ability to make predictions outside the fitting data set is questionable. Furthermore, the recent developments of new generation polarizable force fields\cite{AMOEBA+1,AMOEBA+2,PaesaniJCP2016MBpol,paesanireview,melcr2019accurate} and machine learning (ML) potentials\cite{paesanireview,BehlerNN,CarPhysRevLett.120.143001,smith2017ani,BehlerLAMMPS} have opened perspectives for atomistic simulations of condensed matter systems. These approaches enable high fidelity modeling of the Born-Oppenheimer (BO) energy, and reproduce advanced quantum chemical calculations at a fraction of their computational cost. When reaching such precision on the BO energy, it becomes crucial to account for NQEs explicitly to accurately reproduce experimental observation and take full advantage of the high accuracy achieved\cite{Xantheas2008,XantheasPaesani,pereyaslavetsPNAS2018_NQE_bio, cheng_PNAS2019_water_NN_PI}.

The conceptual and computational complexity of the methods that account for NQEs explicitly has hindered their spread to a broad community. Reliable results can be obtained in the imaginary-time path integrals (PI) framework \cite{feynman2010quantum, chandlerJCP1981PI}, by simulating multiple classical replicas of the system (also called beads). PI provides a numerically exact reference for static properties (approximations have also been derived for dynamical observables, as discussed below), but their numerical cost increases linearly with the number of replicas and can become very large compared to classical molecular dynamics (MD). Several solutions have been proposed to mitigate this cost, such as multiple timestepping in real and imaginary time\cite{markland2008_RP_contraction, cheng2016PI_imaginary_multi_timestepping, kapil2016_RP_contraction_DFT, marsalek2016ab_initio_RP_contraction}. However, this method is based on a decomposition of the energy as a sum of cheap high-frequency and expensive low-frequency term, that is not always feasible (in particular in ML approaches). Other developments, such as high-order PI \cite{perez2011high_order_PI, kapil2016high_order_PI} or PI perturbation theory\cite{poltavsky2016perturbedPI,poltavsky2020high_order_PI_perturbation} allow decreasing the number of necessary replicas, but the computational overhead remains important - typically increasing the simulation load by an order of magnitude for hydrogen-bonded systems at room temperature. 

Recently, a different approach was introduced for the explicit treatment of NQEs with the Quantum Thermal Bath (QTB) \cite{dammak_quantum_2009, Bronstein_PRB2014} and the related quantum thermostat\cite{Ceriotti_PRL2009, ceriotti2010colored_thermostat}, relying on generalized Langevin thermostats to approximate the zero-point motion of the nuclei. Although elegant and inexpensive, these methods suffer from zero-point energy (ZPE) leakage from high to low frequency modes which can lead to massive errors\cite{hernandez-rojas_applicability_2015, brieuc2016_ZPEL}. One possible workaround is to combine the generalized thermostat approach with path integrals\cite{brieuc_JCTC2016_QTB_PI, ceriotti_JCP2011_PIGLET}. Even though the number of required replicas is reduced compared to standard PIMD simulations, the computational cost remains significant (at least 6 beads are needed for water at ambient conditions\cite{Ceriotti2016water_review}). In this letter, we focus on an alternative approach, the adaptive QTB (adQTB) that completely avoids resorting to PI.

In adQTB, the ZPE leakage is compensated directly, using a quantitative criterion derived from the fluctuation-dissipation theorem (FDT). The method was successfully tested on model systems\cite{mangaud_jctc2019}, but its applicability to more realistic problems remained to be demonstrated. In the following, we report the main theoretical aspects of the QTB and adQTB methodologies and introduce two refinements to the adQTB algorithm, improving its efficiency and accuracy and broadening the range of its possible applications, in particular enabling reliable constant pressure simulations. We then apply the method to liquid water. Careful comparison with PI references for structural and thermodynamic properties as well as infrared absorption spectra (IRS) shows that, contrary to standard QTB which is plagued by massive ZPE leakage, adQTB is able to capture NQEs with a remarkable accuracy, while its computational overhead remains limited to less than 25\% compared to classical MD, allowing to scale up the system size to over a million atoms.

In (ad)QTB simulations, each nuclear degree of freedom follows a Langevin equation \cite{dammak_quantum_2009}: 
\begin{equation}
\label{eq_langevin}
    m_i\frac{d^2{r_i}}{dt^2}(t)=-\frac{\partial V}{\partial {r_i}}-m_i\gamma\frac{d{r_i}}{dt}(t)+{R_i(t)} \hspace{0.5cm} i=1 , ..., 3N
\end{equation}
where $V(r_1,...,r_i,...,r_{3N})$ is the interatomic potential ($i$ denotes both the atom number and the direction $x$, $y$ or $z$). 
Eq.~\eqref{eq_langevin} comprises a dissipative force (with friction coefficient $\gamma$) balanced by a random force $R_i(t)$ that injects energy in the system. In classical Langevin dynamics, $R_i(t)$ is a white noise, whose amplitude is proportional to temperature. In QTB, the random force is colored with the following correlation spectrum:
\begin{equation}
    C_{R_iR_j}(\omega)=2m_i\gamma_i(\omega)\theta(\omega,T) \delta_i^j
\end{equation}
where $\gamma_i(\omega)$ is the random force amplitude and
\begin{equation}
    \theta(\omega,T)=\hbar\omega\left[\frac{1}{2}+\frac{1}{e^{\,\hbar\omega/k_BT}-1}\right]
\end{equation}
corresponds to the average thermal energy in a quantum harmonic oscillator at frequency $\omega$ and temperature $T$. Therefore, the aim of the QTB is to account for ZPE contributions in an otherwise classical dynamics by thermalizing each vibrational mode with an effective energy $\theta(\omega,T)$ instead of the classical thermal energy $k_BT$. However, in the original formulation of the QTB (where $\gamma_i(\omega)=\gamma, \forall \omega$), the ZPE provided to high-frequency modes leaks towards low frequencies, which leads to an incorrect energy distribution and can dramatically alter the results. In adQTB, this leakage is quantified precisely using a general result of linear response theory: the quantum fluctuation-dissipation theorem (FDT)\cite{kubo1966fluctuation}.  
For each degree of freedom $i$, we define the deviation from the FDT as:
\begin{equation}
    \label{DeltaFDT_equation}
    \Delta_{FDT,i}(\omega)= \text{Re}\left[C_{v_iR_i}(\omega)\right]-m_i\gamma_i(\omega) C_{v_iv_i}(\omega)
\end{equation}
$v_i=\frac{dr_i}{dt}$ denotes the velocity, while $C_{v_iv_i}(\omega)$ and $C_{v_iR_i}(\omega)$ are respectively its autocorrelation and its cross-correlation spectrum with the random force $R_i$. The FDT characterizes the frequency-dependent distribution of energy in a quantum system at thermal equilibrium. It implies that $\Delta_{FDT,i}(\omega)$ should be zero for any $\omega$, a condition violated in standard QTB, due to ZPE leakage. In adQTB, $\Delta_{FDT,i}(\omega)$ is estimated at regular intervals and the coefficients $\gamma_i(\omega)$ are adjusted on the fly via a first-order dynamics to correct for this violation: a negative $\Delta_{FDT,i}(\omega)$ reveals an excess of energy at frequency $\omega$, so $\gamma_i(\omega)$ is reduced, and conversely for positive deviations. The adQTB results are produced once the $\gamma_i(\omega)$ are adapted and $\Delta_{FDT,i}(\omega)$ vanishes on average.

Here, we introduce two refinements with respect to Ref.~\onlinecite{mangaud_jctc2019}, both of which are presented in full detail in Supplementary Materials. First, to improve the adaptation efficiency, the coefficients $\gamma_i(\omega)$ are adjusted according to the mean FDT deviation, averaged over all equivalent degrees of freedom (i.e. over the 3 directions and over all same-type atoms). Second, we account for the fact that, due to the spectral broadening induced by the friction force, the QTB (and adQTB) tends to slightly underestimate the average potential energy and to overestimate the kinetic energy. This error (unrelated to ZPE leakage) can be predicted and quantified for a harmonic oscillator\cite{barrat_portable_2011,Basire_PCCP2013_QTB_Wigner}. We use this harmonic reference and the deconvolution procedure of Ref.~\onlinecite{rossi_ceriotti_2018_deconvolution} to correct for this inaccuracy: we slightly modify $\theta(\omega,T)$ to compensate for the effect of $\gamma$ on the potential energy, while the kinetic energy is corrected \textit{a posteriori}. The kinetic energy correction is significant (more than 10\%) and essential to enable reliable isobaric simulations, as its neglect causes large errors on the pressure estimation. 

The role of NQEs in liquid water has been extensively investigated both experimentally and theoretically\cite{PhysRevLett.101.017801,doi:10.1021/jp810590c,REITER2004,doi:10.1063/1.1319614,romanelli_JPCL2013_proton_momentum_distribution}. It also represents a major challenge for the adQTB, as massive ZPE leakage takes place from the high-frequency intramolecular vibrations (O-H stretching and H-O-H bending modes) toward the slow intermolecular motion\cite{hernandez-rojas_applicability_2015, habershon_manolopoulos_JCP2009_ZPEL_RPMD_IVR}. Moreover, net NQEs on the structural properties of water are relatively weak due to the competition between two opposite trends: the stretching ZPE strengthens hydrogen bonding, while the bending ZPE weakens it\cite{li2011quantum_H_bond, Ceriotti2016water_review}. The ability of the adQTB to capture this subtle balance is an important indication of its robustness that opens perspectives for its broader application.

\begin{figure*}
    \includegraphics[width=0.95\linewidth]{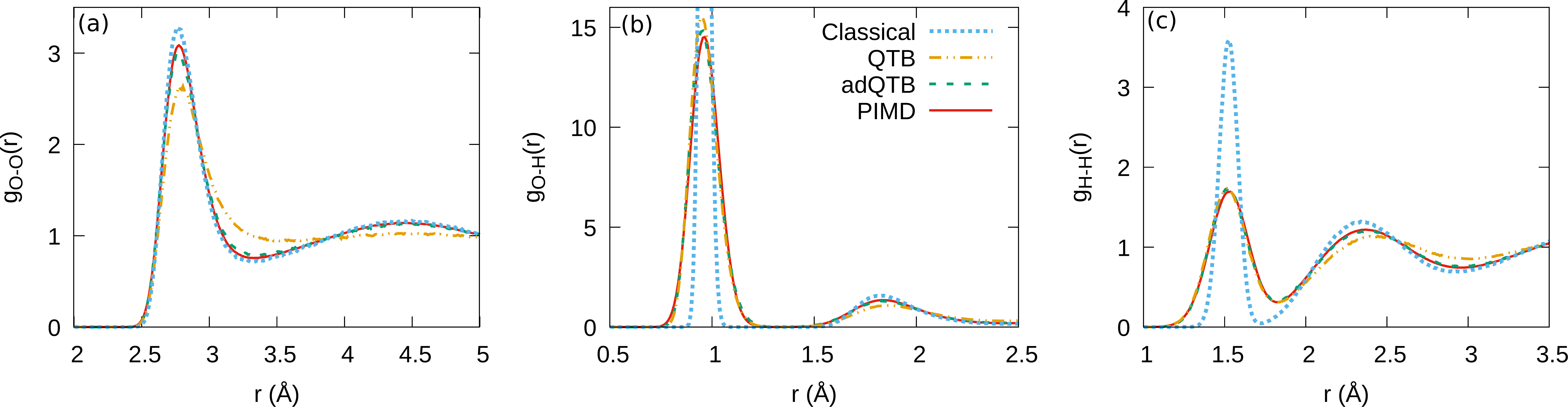}
    \centering
    \caption{\label{RDF} Radial distribution functions at 300~K and constant volume corresponding to the density $\rho=0.995$ g.cm$^{-3}$. }
\end{figure*}

Interatomic interactions are modeled by the q-TIP4P/F potential \cite{Habershon_Manolopoulos_JCP2009_qtip4p_competing} which was included in a local version of the Tinker-HP massively parallel package\cite{tinkerhp}, where we also implemented PIMD and (ad)QTB. Simulations are performed with 1000 water molecules. PIMD simulations are essentially converged with 32 beads (the number typically reported in the literature) and require short timesteps, we used a 0.2~fs timestep for all methods and checked that increasing it to 1~fs had only a limited effect on the accuracy of the adQTB results. In classical Langevin MD and PIMD simulations, static averages are independent of the parameter $\gamma$, and we use $\gamma=1$~ps$^{-1}$ in both cases to limit its effect on dynamical properties. On the other hand, adQTB requires relatively large friction coefficients $\gamma$, to prevent vanishing of $\gamma_i(\omega)$ during adaptation (which whould results in incorrect compensation of the ZPE leakage\cite{mangaud_jctc2019}). We use $\gamma=20$~ps$^{-1}$ for all QTB and adQTB simulation (the influence of these parameters and the scalabilty of the algorithm for large systems is assessed in Supplementary Material).

In Figure~\ref{RDF}, the QTB and adQTB Radial Distribution Functions (RDFs) are compared with their classical and PIMD counterparts. The most salient NQE for this observable is the strong broadening of the intramolecular peaks caused by ZPE in the O-H and H-H RDFs. This effect is very well captured by the adQTB simulations, while it is slightly underestimated by the standard QTB due to ZPE leakage. Apart from this, the classical and quantum RDFs are very similar, due to the aforementioned competition of NQEs. In standard QTB simulations, the leakage of the intramolecular ZPE destabilizes the hydrogen bond network completely and the intermolecular peaks are excessively broadened, but the adQTB procedure efficiently suppresses the leakage and the corresponding curves almost superimpose with the PIMD reference. 


 \begin{table}
    \centering
    $
    \begin{array}{*{12}{c}}
    \hline \hline     
    \text{\small{}} & \text{\small{E$_{k}$}}     &   \text{\small{AB}  }   & \text{\small{BS}}    & \text{\small{VdW}} & \text{\small{Coul.}} & \text{\small{r$_{OH}$(\AA)}} & \text{\small{$\theta_{HOH}$(deg)}} \TBstrut  \\
    \hline
    \text{Classical} & \text{\small{2.69}} &    \text{\small{0.41}} &       \text{\small{1.18}}&       \text{\small{2.20}} &       \text{\small{-14.00}} &       \text{\small{0.96}}  & \text{\small{104.8}}  \TBstrut \\
    \text{QTB} & \text{\small{8.39}} &   \text{\small{1.23}} &       \text{\small{5.81}}&       \text{\small{1.72}} &       \text{\small{-12.38}} &       \text{\small{0.98}}  & \text{\small{104.6}}   \TBstrut \\
    \text{adQTB} & \text{\small{8.60}} &    \text{\small{1.17}} &       \text{\small{6.37}}  &       \text{\small{2.11}}  &       \text{\small{-13.76}}  &       \text{\small{0.98}}  & \text{\small{104.7}}  
    \TBstrut\\
     \text{PIMD} & \text{\small{8.41}} &   \text{\small{1.17}} &       \text{\small{6.26}} &       \text{\small{2.15}} &       \text{\small{-13.87}}  & \text{\small{0.98}}  & \text{\small{104.7}}    \TBstrut \\
    \hline \hline
\end{array}
$
\caption{\label{Observables_values} 
Observables at 300~K. The kinetic energy ($E_k$), the Angular Bending (AB), Bond Stretching (BS), Vand der Waals (VdW) and Coulomb (Coul.) energy terms of the q-TIP4P/F potential\cite{Habershon_Manolopoulos_JCP2009_qtip4p_competing} are reported in kcal.mol$^{-1}$ per water molecule (the standard error is inferior to 0.01~kcal.mol$^{-1}$), together with the average oxygen-hydrogen distance $r_{OH}$ and molecular angle $\theta_{HOH}$.} 
\end{table}

\begin{figure*}
    \centering
    \includegraphics[width=0.95\linewidth]{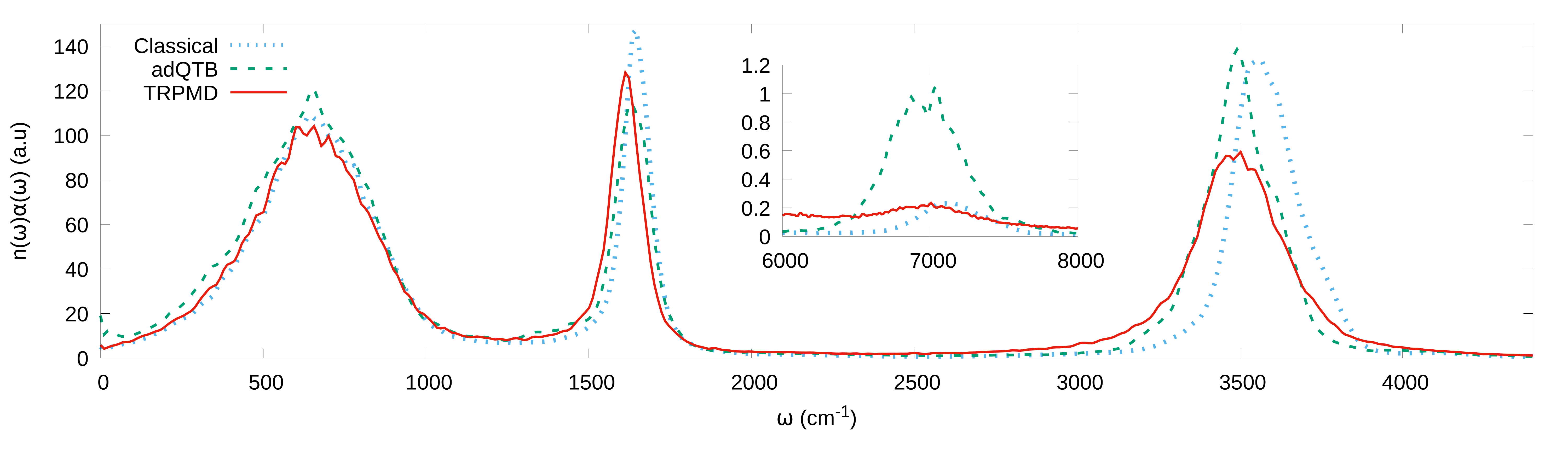}
    \centering
    \caption{\label{IR} Infrared absorption spectra at 300~K (in arbitrary units).} 
    \centering
\end{figure*}

This analysis is further confirmed by Table~\ref{Observables_values}, reporting the average of the different the q-TIP4P/F energy terms. Intermolecular interactions (labeled Coulomb and VdW) are only slightly affected by NQEs and their classical and PIMD values are close. In standard QTB, the total intermolecular energy is overestimated by more than 1~kcal.mol$^{-1}$ due to ZPE leakage, but this is well corrected in adQTB, where accurate values are recovered. The adQTB is remarkably precise for intramolecular energies (labeled AB and BS) and for the kinetic energy (that comprise large amounts of ZPE). It also captures the elongation of the OH distance induced by NQEs, while the molecular angle is essentially unaffected. The dielectric constant computed from the adQTB simulations at 300~K is 57, in good agreement with our PIMD estimation of 58 and with the value in Ref.~\onlinecite{Habershon_Manolopoulos_JCP2009_qtip4p_competing}, given the relatively large statistical uncertainties.

Although PIMD provides a numerically exact reference for static quantum properties, the computation of dynamical observables, such as infrared absorption spectra (IRS), represents a much steeper theoretical challenge, subject of intense research\cite{rossi_ceriottiJCP2014_TRPMD, heleJCP2015matsubara, beutier2015_spectra, ceottoPRL2017_divide_conquer, basire2017fermi,treninsJCP2019_QCMD, ple_JCP2019_Wigner}. There is no reference method to compute IRS exactly while accounting for NQEs in large systems, but various approximations have been developed\cite{Cao_Voth1993,Cao_Voth1994, miller2001semiclassical,Craig_manolopoulosJCP2004}. Recently, Benson et al. compared different state-of-the-art approximate methods for IRS calculation in liquid water and ice\cite{benson2019quantum}. They show that the Linearized semiclassical initial value representation (LSC-IVR) method\cite{miller2001semiclassical} - where time-correlation functions are computed from short classical trajectories initialized from an approximate sampling of the Wigner distribution - provides the most accurate IRS within their broad set of approaches, while the PI-based thermostated ring-polymer MD (TRPMD)\cite{rossi_ceriottiJCP2014_TRPMD} is presented as the cheapest available approach yielding reliable results. QTB has formerly been used with some success as an empirical method to compute approximate IRS\cite{Bronstein_PRB2014, Bronstein_PRB2016}. Although not formally derivable from first principles except for the harmonic oscillator case, the use of QTB and adQTB for IRS calculations can be justified qualitatively by noting that the short-time dynamics is only little affected by the thermostat and thus essentially classical. Therefore, much like LSC-IVR, the QTB combines classical dynamics with approximate quantum initial value sampling. Furthermore, the deconvolution procedure of Ref.~\onlinecite{rossi_ceriotti_2018_deconvolution} efficiently eliminates the main effect of the thermostat: the broadening of the spectral peaks.

Figure~\ref{IR} compares IRS computed in adQTB to those obtained in classical MD and TRPMD (for which a mild Langevin thermostat with $\gamma=1$~ps$^{-1}$ was applied). Compared to TRPMD, the low-frequency absorption band computed with adQTB is slightly more intense, and the bending peak (around 1500~cm$^{-1}$) is a little blue-shifted and broadened. The OH stretching peak at 3500~cm$^{-1}$ is sharper in adQTB than in TRPMD and its overtone at 7000~cm$^{-1}$ has a much larger intensity. These two discrepancies are in favor of the adQTB approach since TRPMD has been shown to cause a spurious broadening of the spectral features and to strongly underestimate anharmonic resonances\cite{benson2019quantum}. Overall, the adQTB IRS are very similar to the LSC-IVR results reported in Ref.~\onlinecite{benson2019quantum}. This should be further confirmed by studies on different systems but it is extremely promising given the almost classical computational cost of adQTB. 

The dynamical properties related to slow molecular motions, on the other hand, cannot be quantitatively assessed in our present adQTB implementation, due to the need for relatively large friction coefficients. The diffusion coefficient $D\simeq 0.8$~cm$^{2}$s$^{-1}$ is underestimated by almost a factor 3 with respect to its RPMD value\cite{Habershon_Manolopoulos_JCP2009_qtip4p_competing} (a similar decrease of $D$ is observed in classical Langevin MD using $\gamma=20$~ps$^{-1}$). The deconvolution procedure is of no help here, since $D$ corresponds to the zero-frequency component of the vibration spectrum, and the deconvolution does not provide reliable results in that spectral region\cite{rossi_ceriotti_2018_deconvolution}. Improved diffusion estimates might be obtained in future works by decreasing $\gamma$ selectively at low frequencies using a generalized friction force, or by appropriately redesigning the adQTB algorithm, for example using the recently introduced fast-forward Langevin method\cite{Hijazi_ceriotti_JCP2018_fast_forward_langevin}. 

\begin{figure}
    \centering
    \includegraphics[width=0.90\linewidth]{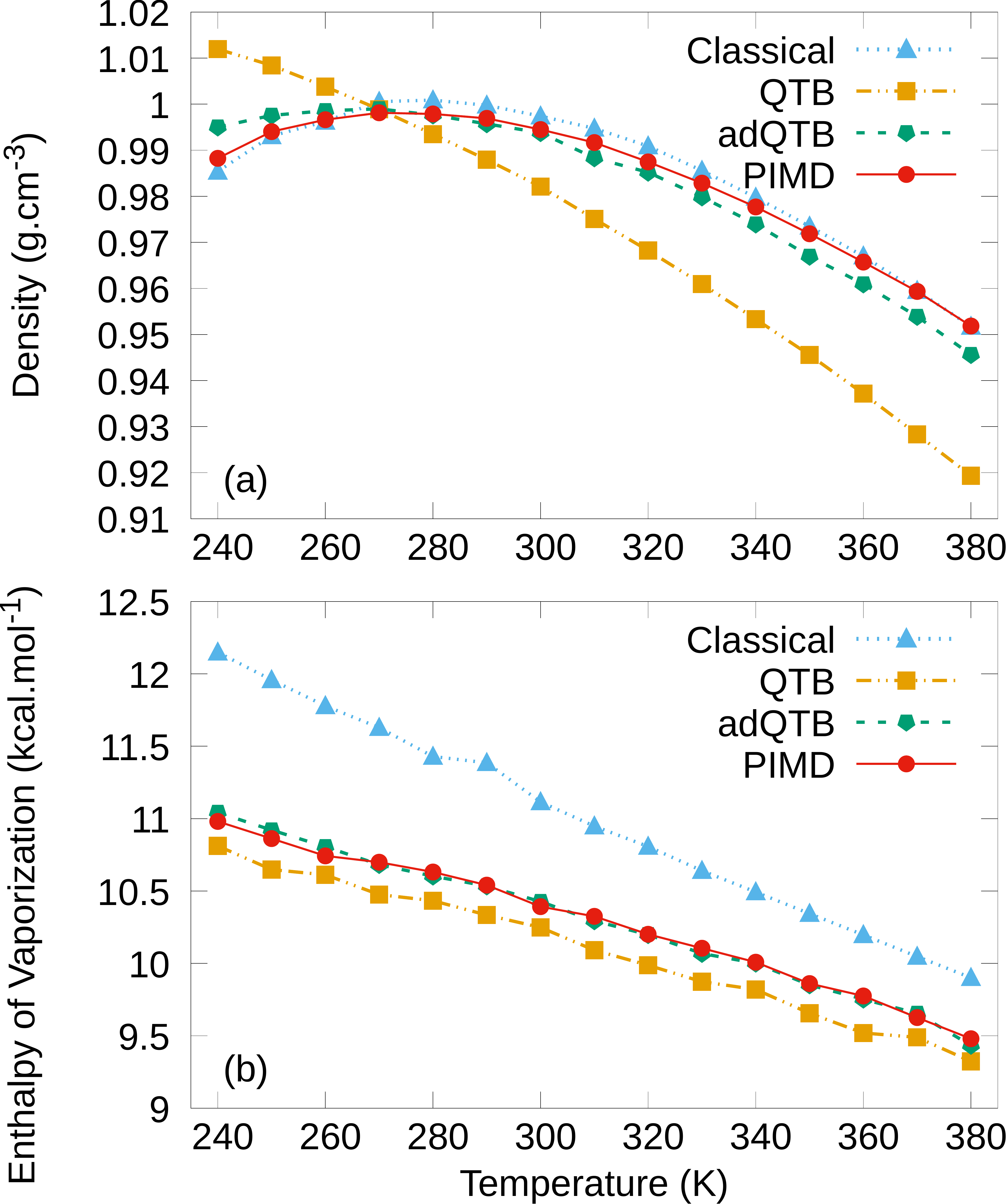}
    \centering
    \caption{\label{Density} Density of liquid water (a) and Enthalpy of vaporization (b) at constant pressure $P=1$~atm.}
\end{figure}

We now explore the use of adQTB to perform fixed pressure simulations using a Langevin piston barostat\cite{feller1995constant,ceriotti2014ipi}. Pressure is a challenging quantity to evaluate in the (ad)QTB framework: its estimator is a difference between two large terms that almost cancel (a potential and a kinetic term, of the order of 10$^5$~atm each). Therefore, even small inaccuracies on either of these contributions can result in non-negligible errors (see Supplementary materials). The results obtained for the density as a function of temperature at $P=1$~atm are shown in Figure \ref{Density}. 

Because of the competition between NQEs, the classical and PIMD results are very similar, both showing a characteristic bell shape with a maximum around 280~K. NQEs are only responsible for a small decrease of the density in the intermediate temperature range (270-330~K). The standard QTB completely fails to capture this temperature-dependence. It decreases monotonously and strongly overestimates the variations of the density. Compensating the leakage in adQTB allows recovering the overall bell shape and a good agreement with the PIMD reference. In the intermediate temperature range (most relevant for biological systems), adQTB is very accurate. The curvature of the density curve is only slightly underestimated, leading to small errors of the order of $0.005$~g.cm$^{-3}$ in the low-temperature and in the high-temperature limits. Note that in barostated simulations, ZPE leakage can take place from the atomic system towards the fictitious piston degree of freedom, but this leakage can easily be avoided by an appropriate choice for the piston mass and friction parameters (see Supplementary materials).

These results show that adQTB can be a useful and inexpensive tool for constant pressure simulations of physical and chemical properties. As an  illustration, we present on Figure~\ref{Density}.b the enthalpy of vaporization $\Delta H_{vap}$ computed from the same isobaric simulations. The classical $\Delta H_{vap}$ is systematically overestimated compared to the corresponding PIMD values\cite{PaesaniJCP2016MBpol, Guillot_JCP1998_NQE_water_Feynman_Hibbs}. When NQEs are included with the standard QTB, $\Delta H_{vap}$ decreases markedly, and becomes even underestimated, but this is due to ZPE leakage and the adQTB recovers an almost perfect agreement with the PIMD reference. 

Finally, we discuss the computational overhead of the adQTB simulations with respect to classical Langevin MD. A first additional cost comes from the generation of the colored random forces and the adaptation of the $\gamma_i(\omega)$ coefficients. It represents approximately 20\% of the total simulation time and the scalability tests provided in Supplementary Materials show that, even for systems over one million atoms, it remains inferior to 25\% in our present implementation - that will be further accelerated using Graphics Processing Units (GPUs)\cite{Tinker-HP-GPUs}. The q-TIP4P/F water model is particularly inexpensive, and we expect this overhead to become negligible in comparison to atomic force calculations with more realistic models. A second additional cost comes from the adaptation procedure that requires time for the $\gamma_i(\omega)$ to converge. This necessary time can vary from one system to another. In our liquid water simulations, we show in Supplementary Materials that with an appropriate choice of adaptation parameters, the $\gamma_i(\omega)$ coefficients can converge in about 10~ps. The minimum adaptation time is thus small compared with the several ns required to reach statistical convergence on some of the physical observables, as the density and the dielectric constant.

The adQTB renews the original promise of the QTB method to provide approximate quantum simulations at an almost classical cost, but with a much improved reliability. It is a promising alternative to PI methods to account for NQEs explicitely in the calculation of static properties as well as vibrational spectra. Combined with accurate ML potentials or polarizable force fields, it should provide a powerful tool with broad applications, in particular for the large-scale simulations required in biophysics and biochemistry. 

\section*{Acknowledgements}
This work was made possible thanks to funding from the European Research Council (ERC) under the European Union's Horizon 2020 research and innovation program (grant agreement No 810367), project EMC2. Computations have been performed at CINES on the Occigen machine on grant no A0070707671. The authors are grateful to Fabio Finocchi and Philippe Depondt for many interesting discussions.
\\

\noindent * louis.lagardere@sorbonne-universite.fr,\\
* jean-philip.piquemal@sorbonne-universite.fr,\\
* simon.huppert@sorbonne-universite.fr


\bibliography{biblio.bib}

\end{document}